\documentclass{egpubl}
 
\JournalSubmission    
\usepackage[T1]{fontenc}
\usepackage{dfadobe}  

\usepackage{cite}  
\BibtexOrBiblatex
\electronicVersion
\PrintedOrElectronic

\ifpdf \usepackage[pdftex]{graphicx} \pdfcompresslevel=9
\else \usepackage[dvips]{graphicx} \fi

\usepackage{egweblnk} 

\usepackage{xcolor}
\usepackage{amsmath}
\usepackage{mathtools}
\usepackage{amsfonts}
\usepackage{booktabs}
\usepackage{multirow}

\newif\ifCommentsAuthors

\CommentsAuthorstrue
\ifCommentsAuthors

    \definecolor{myred}{rgb}{.8,.0,.0}
    
    \definecolor{myblue}{rgb}{0,0.4,.8}
    \definecolor{mygreen1}{rgb}{0.1,0.9,0}
    \definecolor{mygreen2}{rgb}{0,0.7,0.3}
    \newcommand{\commentmarc}[1]{\textcolor{mygreen1}{>>#1<<}}
    \definecolor{myblue2}{rgb}{0.1,0,.9}
    
    \definecolor{mcolor}{rgb}{0.8,0.0,0.3}
    
    \definecolor{mcolor2}{rgb}{0.7,0.0,0.5}
    
    \definecolor{saeedcolor}{HTML}{FF6600}
    \newcommand{\commentsaeed}[1]{\textcolor{saeedcolor}{>>#1<<}}
    \definecolor{saeedcolor2}{HTML}{b942f5}

    \definecolor{mypink}{rgb}{1.0,.5,0.5}
    \newcommand{\commentdan}[1]{\textcolor{mypink}{#1}}
    
\else
    \definecolor{myred}{rgb}{.8,.0,.0}
    
    \newcommand{\commentmarc}[1]{}
    \newcommand{\commentsaeed}[1]{}
    \newcommand{\commentdan}[1]{}
    
\fi

\newif\ifSubmission
\newcommand{\submissionID}{6033}
\Submissionfalse

\usepackage{subcaption}
\usepackage{graphicx}

\title[ZEGGS]%
      {ZeroEGGS: Zero-shot Example-based Gesture Generation from Speech}

\ifSubmission
\author[\submissionID]
{\parbox{\textwidth}{\centering Anonymous}
{\parbox{\textwidth}{\centering}
}
}
\else
\author[S. Ghorbani, Y. Ferstl, D. Holden, N.F. Troje, M. Carbonneau]
{\parbox{\textwidth}{\centering 
S. Ghorbani$^{1,2}$\orcid{0000-0002-3227-9013}, 
Y. Ferstl$^{1}$\orcid{0000-0001-7259-0378}, 
D. Holden$^{1}$\orcid{0000-0000-0000-0000}, 
N. F. Troje$^{2}$\orcid{0000-0002-1533-2847}
M. Carbonneau$^{1}$\orcid{0000-0002-0677-415X}
}
        \\
{\parbox{\textwidth}{\centering $^1$Ubisoft, Canada\\
         $^2$York University, Canada
       }
}
}
\fi

\begin{document}


\teaser{
 \includegraphics[width=\linewidth]{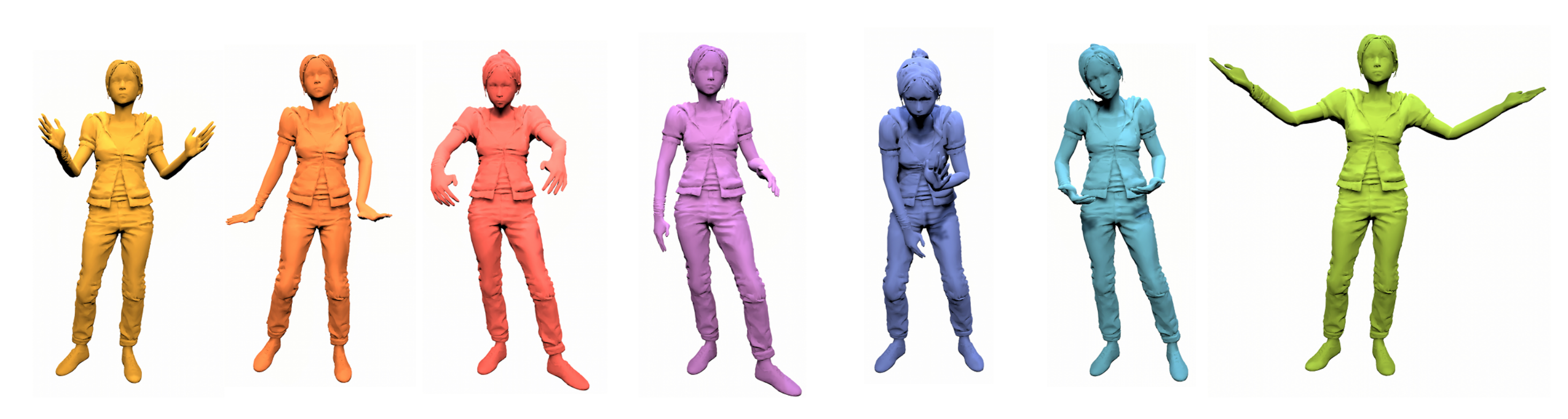}
 \centering
  \caption{Samples generated by our model in different styles. Left to right: \textit{Neutral}, \textit{Happy}, \textit{Angry}, \textit{Relaxed}, \textit{Old}, \textit{Sad}, \textit{Oration}.}
\label{fig:teaser}
}
\maketitle
\begin{abstract}
We present ZeroEGGS, a neural network framework for speech-driven gesture generation with zero-shot style control by example. This means style can be controlled via only a short example motion clip, even for motion styles unseen during training. Our model uses a Variational framework to learn a style embedding, making it easy to modify style through latent space manipulation or blending and scaling of style embeddings. The probabilistic nature of our framework further enables the generation of a variety of outputs given the same input, addressing the stochastic nature of gesture motion. In a series of experiments, we first demonstrate the flexibility and generalizability of our model to new speakers and styles. In a user study, we then show that our model outperforms previous state-of-the-art techniques in naturalness of motion, appropriateness for speech, and style portrayal. Finally, we release a high-quality dataset of full-body gesture motion including fingers, with speech, spanning across 19 different styles. Our code is publicly available at \url{https://github.com/ubisoft/ubisoft-laforge-ZeroEGGS}.


\end{abstract}  
\section{Introduction}

In human communication, gestures complement speech by providing additional information about thoughts, feelings, emotions, and intentions \cite{melinger2004gesture,de2012interplay}. Efficiently animating realistic gesture behavior is a key concern in many applications based on virtual humans, such as characters in video games and extended realities or customer service agents.

A common method for the automation of gesture animation is to trigger pre-recorded animations from a database based on tags. One can manually mark-up a dialog or program systemic events. While effective, this method requires significant amounts of time and manpower in order to produce a variety of engaging and realistic animations. Therefore this method does not scale up well with increasingly large open world video games and cannot be applied to avatars in virtual worlds where dialog content is not scripted. Moreover, pre-recorded animation may not be synchronized with speech rhythm.

These problems of scaling and synchronicity have motivated research into methods for automatic generation of gestures with data-driven methods \cite{alexanderson2020style,yoon2020speech}. Nonetheless, despite recent research efforts, generating realistic gesture motion remains a difficult problem, and addressing expressivity of speaker state and identity by providing control over style is even more challenging. 

Speech is commonly used as a control input for gesture generation systems. Yet it is a weak control signal because a speech line may be associated with many different motions. This means that there is not only one true gesture match, but many equally appropriate gestures. In other words this is a one-to-many mapping problem. Moreover, every person exhibits their own style of movement, and therefore, producing relatable and engaging stories with variety of characters requires the ability to generate such distinct and appropriate styles. This implies that the model must capture this wide range motion variation, preserving characters' idiosyncrasies.

Previous works have sought to address the problem of creating distinct styles by modelling and generating gestures for specific speakers \cite{neff2008gesture,ginosar2019learning,yoon2020speech,ahuja2020style} and by modifying gesture motion through general statistics such as hand height and velocity \cite{alexanderson2020style,yoon2021sgtoolkit}. These approaches lack flexibility because they are limited by the content of the training data. They require examples of every target speakers and every style prior to training the model, and cannot generalize outside of this range. This means a specific dataset needs to be captured for each individual and each style, which leads to a prohibitive amount of work in larger scale applications. To scale up, a gesture generation method must be able to capture an individual style with a very limited amount of data, ideally only one example. 
Secondly, motion style can be difficult to capture in words, rather, it can be easier to describe style by providing an example of the desired motion style.

In this work, we propose ZeroEGGS, a system that generates stylized full body gestures from speech. Unlike existing solutions, ZeroEGGS encodes gesture style from short example motions, which allows for zero-shot style transfer. This means that it can generalize to styles that were not covered by the training data, and does not require style labels. Moreover, the model learns a meaningful representation that enables style manipulation directly in the latent space. Our framework is probabilistic, allowing for repeated sampling to obtain a variety of output motion given the same input speech and style. Finally, we release a high-quality multi-modal dataset of synchronized speech and full-body gesture in 19 different styles. We also make our code publicly available for reproducibility\footnote{\url{https://github.com/ubisoft/ubisoft-laforge-ZeroEGGS}}. 

\section{Related Work}

Research addressing automatic production of gesture motion from speech can be divided into two main approaches: rule-based, and data-driven methods.

Rule-based methods use explicit, often manually created, mappings of speech markers to gestures. Some methods rely on text analysis \cite{Cassell2001, Lee2006, kopp2003max}, while others use acoustic features \cite{Marsella2013}. Rule-based approaches provide author-control, high motion definition and facilitate the generation of semantic gestures. However, motion diversity is limited by the number of designed rules and database content. Moreover, the required manual labor hinders scalability.

Data-driven methods model correlations between speech and motion features rather than relying on hand-crafted mappings. Some methods still require manual labour due to using hand-annotated features such as the gesture shape \cite{neff2008gesture} or timing of stressed phrases \cite{yang2020}. The mapping between the motion and speech features is learned automatically. At inference, motion is produced by concatenating snippets from a database, which retains captured motion quality. Other approaches generate motion instead of drawing from a database. A model must learn a mapping between speech and gesture, which is a one-to-many problem. In that situation, minimizing the error between predicted and target motion often leads to mean collapsing problems and lethargic motion with small ranges (e.g. \cite{ferstl2018investigating} and \cite{kucherenko2020gesticulator}). Generative adversarial networks have been used to address this problem with varying degree of success depending on the datasets size and gesture complexity \cite{ferstl2019multi, ginosar2019learning,rebol2021passing}.

Despite recent research efforts, the naturalness of generated gestures still falls significantly behind motion-captured gestures, and often fails to outperform even mismatched real motion w.r.t. appropriateness \cite{kucherenko2021large,ginosar2019learning}.

While the aforementioned methods condition generation only on speech, several recent efforts propose to also provide control over gesture style. 
In \cite{alexanderson2020style}, style control is achieved via input of four desired motion statistics, specifically the gesture speed, height, spacial extent and lateral symmetry. The authors employ a probabilistic model that predicts the next pose distribution instead of predicting a fixed pose; gesture motion can then be re-sampled repeatedly to obtain a variety of sequences. 
Similarly, in \cite{yoon2021sgtoolkit} a gesture generation toolkit is presented with the control parameters speed, spacial extent, and handedness.  The system in \cite{sonlu2021conversational} uses the Laban Effort and Shape qualities as animation modifiers to impart the intended personality to the character. All these methods rely on handcrafted control features which are often not descriptive enough to encode to wide range of distinct styles that can be learned by the model. 

It was proposed to use style examples to address the shortcomings of hand-crafted features for other animation applications. For instance, Aberman et al.~\cite{aberman2020unpaired} transferred style in human locomotion using example clips. While the method can generalize to unseen styles, providing a large enough dataset, it requires a set of labeled styles to train the discriminator in the model. Similarly, Valle-Pérez et al.~\cite{valle2021transflower} introduced a model for dance motion generation conditioned on music and a short style example motion. However, the model was not shown to generalize beyond motion styles contained in the training data.

The inherent difficulty of accurately naming styles and collecting meaningful style datasets is not unique to the animation domain. For instance, recent speech synthesis efforts focused on speech stylization by example and enable style generalization beyond the training data. 
Some works \cite{Hsu2018, Wang2018, zaidi2021daft} augment a traditional text-to-speech model with a style encoder capturing the general style and prosody of a line to condition the generation. These models can generalize to new styles, but also allow for interpolation in the style latent space. 
Other work has even made style generalization for unseen speakers possible \cite{zaidi2021daft}.
We take inspiration from these methods and adapt their ideas to gesture generation to inherit their advantages.

\section{System Overview}

\begin{figure}[t]
  \centering
  \includegraphics[width=1.2\linewidth]{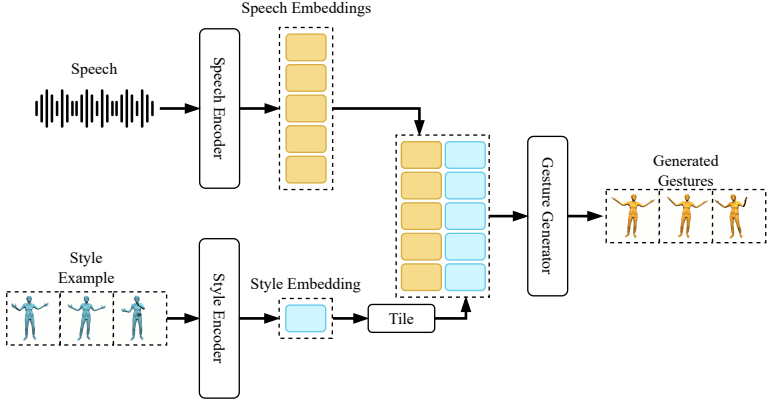}
  \caption{An overview of proposed model.}
  \label{fig:overview}
\end{figure}

Fig. \ref{fig:overview} shows an overview of the ZeroEGGS architecture which can be divided into three components: (1) Speech Encoder, (2) Style Encoder and (3) Gesture Generator. The Speech Encoder translates a sequence of raw speech data into a speech embedding sequence $S$. The Style Encoder summarizes the style of a reference animation sequence into a fixed-size style embedding vector $\textbf{e}$. Finally, an autoregressive decoder called Gesture Generator uses the style embedding vector concatenated to the speech embedding sequence to generate the corresponding gesture animation $Y$. This model is trained by maximizing the likelihood of the gesture animation $p(Y \mid S, e) = \prod_{t} p\left(\mathbf{y}_{t} \mid \mathbf{y}_{<t}, S, \mathbf{e}\right)$.

\subsection{Speech Encoder}

\begin{figure}[t]
     \centering
     \begin{subfigure}[b]{0.15\textwidth}
         \centering
         \includegraphics[width=\textwidth]{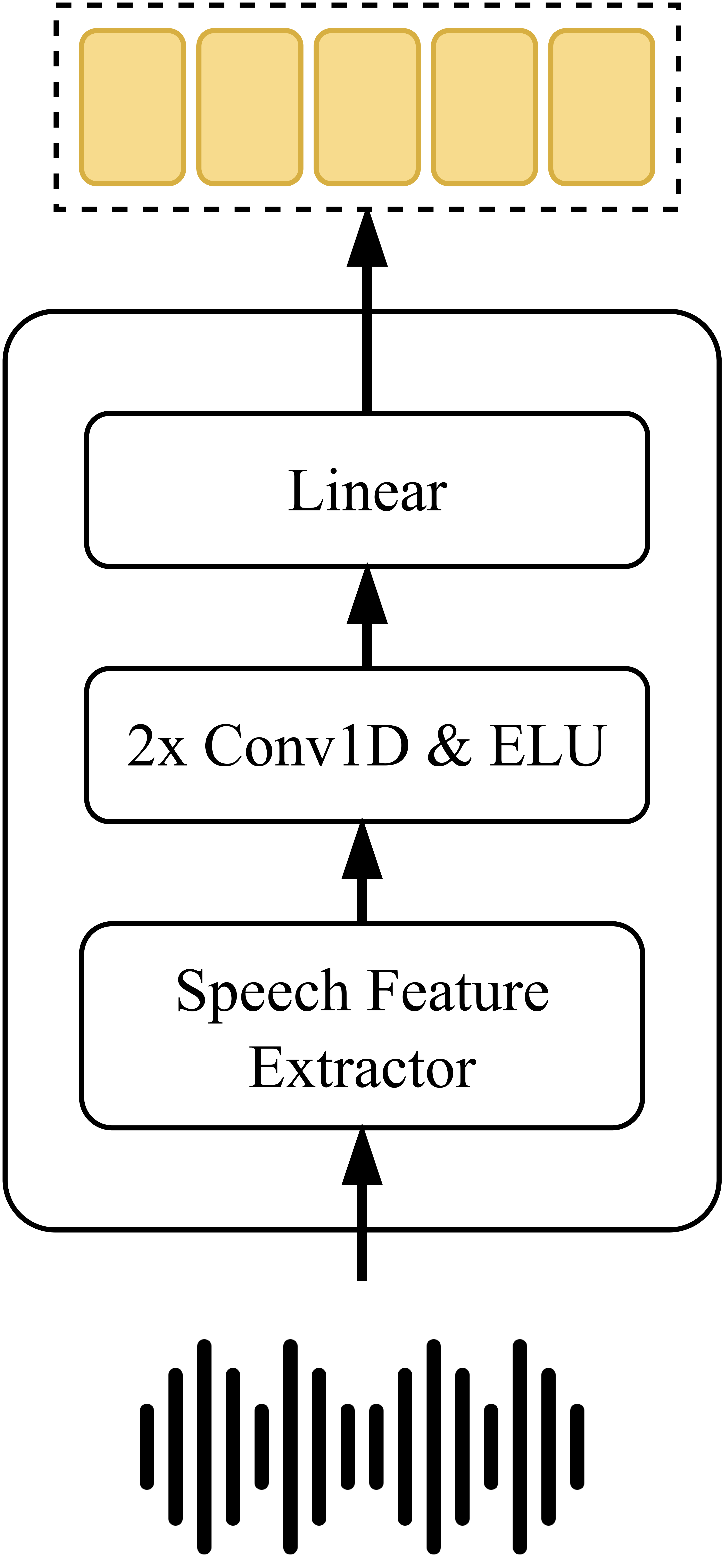}
         \caption{Speech Encoder}
         \label{fig:Speech Encoder}
     \end{subfigure}
        \hspace{.1 in}
     \begin{subfigure}[b]{0.17\textwidth}
         \centering
         \includegraphics[width=\textwidth]{figures/v2/StyleEncoder_v2.pdf}
         \caption{Style Encoder}
         \label{fig:Style Encoder}
     \end{subfigure}
        \caption{Architectures of the Speech Encoder and the Style Encoder}
        \label{fig:Speech and Style encoders}
\end{figure}

The Speech Encoder, illustrated in Fig. \ref{fig:Speech Encoder}, converts raw audio input into a sequence of speech embedding vectors. First, we convert raw audio samples to spectrograms. We use the log-amplitude of the spectrogram and the mel-frequency scale as done in many speech applications \cite{zaidi2021daft, hannun2014deep}. Spectrograms capture how frequency components of a signal vary across time, and the mel scale for frequency better approximates how humans perceive sounds. We also extract the energy per frame as a supplementary feature.
Then, extracted features are re-sampled and passed through 1D convolution layers followed by non-linear operators, and finally a frame-wise linear layer. This results in a sequence of embedding vectors $S = [\textbf{s}_0, \textbf{s}_1, \dots, \textbf{s}_{T-1}] \in \mathbb{R}^{T \times {D}_{S}}$ where $T$ is the number of frames in the sequence and ${D}_{S}$ is the size of the speech embedding vector for each frame.

\subsection{Style Encoder}
\label{section:style_encoder}
The Style Encoder summarizes the reference style animation clip into a low dimensional, fixed size, embedding vector which captures general attributes of the reference style. The style embedding is sampled from a multivariate Gaussian distribution as described in the Variational Auto-Encoder (VAE) framework \cite{kingma2013auto}. VAE models are known to learn latent spaces that can be disentangled \cite{Rolinek2019} and interpolated \cite{Bowman2016}. Moreover, in our application they naturally allow for the sampling of variations at inference time.

Each frame of the reference animation clip is represented by a feature vector $\mathbf{a}=[\boldsymbol{\rho}_{p}, \boldsymbol{\rho}_{r}, \dot{\boldsymbol{\rho}}_{p}, \dot{\boldsymbol{\rho}}_{r}, \dot{\mathbf{r}}_{p}, \dot{\mathbf{r}}_{r}]$ where $\boldsymbol{\rho}_{p}\in \mathbb{R}^{3j}, \boldsymbol{\rho}_{r}\in \mathbb{R}^{6j}$ are the joint local translations and rotations, $\dot{\boldsymbol{\rho}}_{p} \in \mathbb{R}^{3j}$ and $\dot{\boldsymbol{\rho}}_{r} \in \mathbb{R}^{3j}$, are the joint local translational and rotational velocities, and $\dot{\mathbf{r}}_p \in \mathbb{R}^{3}$ and $\dot{\mathbf{r}}_{r} \in \mathbb{R}^{3}$ are the character root translational and rotational velocity local to the character root transform. $j$ corresponds to the number of joints in the kinematic tree. Joint rotations are represented by 2-axis rotation matrix while joint and root rotational velocities are specified using the scaled angle axis representation as done in \cite{zhang2018mode}. We compute the root position of the character by projecting the position of the second spine joint on the ground. We project the \textit{z-axis} of the hip joint onto the ground to obtain the root orientation.

The sequence of $M$ feature frames, $A = [\textbf{a}_{0}, \textbf{a}_{1}, \dots, \textbf{a}_{M-1}]$, is normalized and fed to a neural network to obtain the style embedding vector $\textbf{e}\in \mathbb{R}^{D_{e}}$, where ${D}_{e}$ is the dimensionality of this final conditioning style embedding vector. The architecture, shown in Fig. \ref{fig:Style Encoder}, is inspired from \cite{zaidi2021daft} which was originally proposed for speech prosody encoding. First, the sequence of animation features, $A$, are passed through two 1D convolution layers each followed by a ReLU and a layer normalization layer. Positional encoding encourages the model to encode the sequence ordering \cite{vaswani2017attention}. Then, similarly to \cite{zaidi2021daft}, we apply a Feed-Forward Transformer block that implements a multi-head self-attention layer \cite{vaswani2017attention} and two 1D convolution layers each followed by a residual connection and layer normalization. This results in a sequence of shape $M \times 2{D}_{e}$. We average over the whole sequence to obtain the parameters $\mathbf{\mu}$ and $\mathbf{\sigma}$ of the ${D}_{e}$-dimensional multivariate Gaussian distribution from which we sample the final style embedding vector $\textbf{e}$.

\subsection{Gesture Generator}\label{section:gesture_generator}

\begin{figure}[t]
  \centering
  \includegraphics[width=0.9\linewidth]{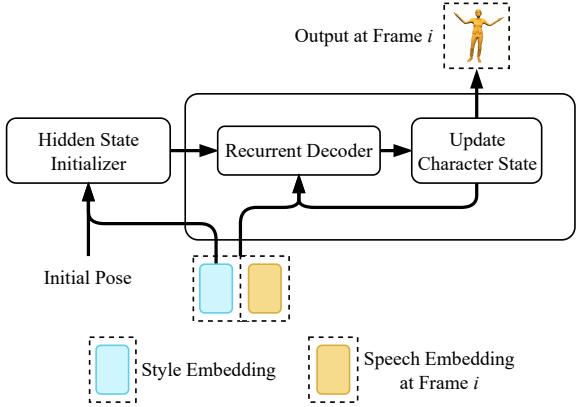}
  \caption{The architecture of Gesture Generator block.}
  \label{fig:gesture_generator}
\end{figure}

The Gesture Generator, shown in Fig. \ref{fig:gesture_generator}, is a conditional auto-regressive model which produces the final animated gesture sequence $Y = \{\mathbf{y}_{0}, \mathbf{y}_{1}, \dots, \mathbf{y}_{T-1}\}$ from the speech embedding sequence $S$ and the reference style embedding vector $\textbf{e}$. The full parametrization of the output pose state for each frame is given by $\mathbf{y} = [\boldsymbol{\rho}_{p}, \boldsymbol{\rho}_{r}, \dot{\boldsymbol{\rho}}_{p}, \dot{\boldsymbol{\rho}}_{r}, {\mathbf{r}}_{p}, {\mathbf{r}}_{r}, \dot{\mathbf{r}}_{p}, \dot{\mathbf{r}}_{r}]$. Similar to style feature representation (see Section \ref{section:style_encoder}), $\boldsymbol{\rho}_{p}, \boldsymbol{\rho}_{r}, \dot{\boldsymbol{\rho}}_{p}, \dot{\boldsymbol{\rho}}_{r}$ are joint local translations and rotations along with their velocities,  and $\dot{\textbf{r}}_{p}$ and $\dot{\textbf{r}}_{r}$ are the character root translational and rotational velocity local to the character root transform. ${\mathbf{r}}_{p} \in \mathbb{R}^{3}$ and ${\mathbf{r}}_{r}\in \mathbb{R}^{4}$ are the position and orientation of the root (represented as quaternions), respectively, which are updated using the root translational and rotational velocities at each frame.

The core of the gesture generator is the Recurrent Decoder. It is an auto-regressive neural network built by two layers of Gated Recurrent Units (GRU). 
It produces the pose encoding for a new frame $i$ from the corresponding speech frame $\mathbf{s}_i$, the reference style embedding vector $\textbf{e}$, and the previous pose state vector $\mathbf{y}_{i-1}$. 

The Update Character State block in Fig. \ref{fig:gesture_generator} formats the Recurrent Decoder output, computes the pose state, and updates the character facing direction. To compute the pose state at each frame, we denormalize the output of the Recurrent Decoder and use the predicted root translational and rotational velocities to update the root transform. In addition to the last pose encoding, we condition the Recurrent Decoder to a fixed target facing direction to avoid rotational drifting over time. We start by converting the target facing direction from world space to the character root transform. Then, we concatenate it with the previously generated pose state, normalize the resulting vector and provide it to the Recurrent Decoder.

The Hidden State Initializer is a separate neural network that provides hidden states for the GRU layers based on the initial pose, the character facing direction and the style embedding. As in \cite{Harvey2018}, we found that using a separate initializing network improved the quality of our results. We implemented the Hidden State Initializer using three linear layers followed by ELU activation functions.
\section{Implementation and Training }
\subsection{Dataset and Data Preparation}
We recorded a high quality dataset of synchronized motion capture and audio. It contains 67 sequences of monologue performed by a female actor speaking in English and covers $19$ different motion styles. The styles were chosen to cover a variety of postures (e.g. \textit{Tired} \textit{vs.} \textit{Oration} in Fig.~\ref{fig:unseen_styles}) as well as hand and head movement (e.g. \textit{Oration}, \textit{Agreement}). The total length of the dataset is 135 minutes. Table~\ref{tab:data} summarizes the information about the captured styles. 

We recorded full-body motion at 60 frames per second (fps) and represented animation data via a skeleton of $j=75$ joints, including hands and fingers. For training, we added the mirrored version of all animation data to double the amount of data and extracted feature vector $\textbf{a}$ and pose state vector $\textbf{y}$ for each frame as explained in sections \ref{section:style_encoder} and \ref{section:gesture_generator}, respectively.
In addition, we extracted the head \textit{z-axis} direction for all frames in the sequence and projected it onto the ground. Then we used the median of the extracted head direction across all frames as the global target facing direction. During runtime, we set the target facing direction to the global \textit{z-axis} direction.

Audio data was recorded at a sampling rate of 48kHz. For the speech encoding, we extract spectrograms using an FFT Hanning-window of 50 ms and a hop length of 12.5 ms. We project the spectrogram into the mel frequency scale, and use the log amplitude of each of the 80 channels as well as the total frame energy as final speech features. We then re-sampled the sequence of speech features to $60$ fps and normalized it. 

\begin{table}[]
\centering
\resizebox{\columnwidth}{!}{
\begin{tabular}{||c|c||c|c||}
\hline
\textbf{Style}        & \textbf{Length (mins)} & \textbf{Style}       & \textbf{Length (mins)}   \\ \hline\hline
Agreement    & 5.25         & Pensive      & 6.21          \\ \hline
Angry        & 7.95         & Relaxed     & 10.81           \\ \hline
Disagreement & 5.33         & Sad         & 11.80          \\ \hline
Distracted   & 5.29         & Sarcastic   & 6.52           \\ \hline
Flirty       & 3.27         & Scared      & 5.58          \\ \hline
Happy        & 10.08         & Sneaky      & 6.27           \\ \hline
Laughing     & 3.85         & Still       & 5.33           \\ \hline
Oration      & 3.98         & Threatening & 5.84           \\ \hline
Neutral      & 11.13        & Tired       & 7.13           \\ \hline
Old          & 11.37         & \textbf{Total}       & \textbf{134.65}          \\ \hline
\end{tabular}}
\caption{Details of the recorded motion and audio dataset in minutes.}
\label{tab:data}
\end{table}

\subsection{Model Implementation}





We adjust the kernels of 1D convolution layers for the speech encoder network to obtain a final receptive field covering roughly 1 second of speech. The first convolutional layer has 64 channels and a kernel size of $3$, the second has 64 channels and a kernel size of $31$. Both convolution layers are followed by a dropout layer with a rate of 0.2 and ELU activations. Our final encoder dimensionality (${D}_{S}$) is 64. 

For the style encoder, we set the dimensionality of the style embedding vector $\mathbf{e}$ to 64. All convolutional layers have a kernel size of 3 with 512 output channels, followed by dropout layers with a drop rate of 0.2. We constructed our Feed-Forward Transformer block with a 4-head self-attention layer, followed by a dropout at a rate of 0.1 and two 1D convolution layers with a kernel size of 3 and 64 channels.

The Recurrent Decoder in our gesture generator is constructed by two GRU cells with a hidden state size of $1024$. The cell state encoder is a 3-layer fully connected feed-forward network with ELU activation and a hidden size of $1024$.
\subsection{Training \& Losses}
We train the network end-to-end using the Rectified Adam optimizer \cite{liu2019variance} with a learning rate of $1e-4$, a decay factor of 0.995 is applied at every 1k iterations. We used a batch size of 32 and stopped training after 120k iterations based on visual result quality. In a batch, the length of the sequences $T$ is set to $256$ frames ($4.26$ seconds).
We do not use teacher forcing during training, but instead train the model on its own predictions. Although this decelerates convergence, it ensures that the model learns to recuperate from its own errors which leads to more robustness. The style example sequence $A$ is sampled from the same animation clip as the target sequence. Its length $M$ randomly spans from 256 to 512 frames, but always encompasses the target sequence. This random sampling scheme ensures de-correlation between styles and clip length. We also randomly alter the animation and speech speed by 10\% in whole batches as a data augmentation strategy.

We can consider our model as a conditional VAE where the objective is to maximize the evidence lower bound (ELBO) of the marginal log likelihood of gesture motion given a speech sequence. Thus we can formulate the training loss as the negative ELBO expressed as
\begin{equation}
\begin{aligned}
\mathcal {L} &= \mathbb{E}_{q\left(\mathbf{z} \mid \mathbf{e}\right)}\left[-\log p\left({Y} \mid {S}, \mathbf{z}\right)\right] + D_{K L}\left(q\left(\mathbf{z} \mid \mathbf{e}\right) \| p\left(\mathbf{z}\right)\right) \\
&= \mathcal{L}_{recon} + D_{K L}\left(q\left(\mathbf{z} \mid \mathbf{e}\right) \| p\left(\mathbf{z}\right)\right)
\end{aligned}
\end{equation}
The first term is hereby the expected negative log-likelihood of the gesture motion, or reconstruction loss. The second term is the regularization term expressed as the Kullback–Leibler divergence between the posterior distribution $q\left(\mathbf{z} \mid \mathbf{e}\right)$ predicted by the style encoder and the prior distribution $p\left(\mathbf{z}\right)$, which is a standard multivariate Gaussian distribution. In our implementation, we used the cost annealing strategy proposed by \cite{Bowman2016} for weighting the regularization term.

Our reconstruction loss divides into several terms:
\begin{equation}
\begin{aligned}
    \mathcal{L}_{recon} =  &\lambda_{p}\mathcal{L}_{p} + \lambda_{r}\mathcal{L}_{r} + \lambda_{vp}\mathcal{L}_{vp} + \lambda_{vr}\mathcal{L}_{vr} + \\ 
    &\lambda_{dp}\mathcal{L}_{dp} + \lambda_{dr}\mathcal{L}_{dr} + \lambda_{f}\mathcal{L}_{f}
\end{aligned}
\end{equation}

where $\mathcal{L}_{p}$, $\mathcal{L}_{r}$, $\mathcal{L}_{vp}$, and $\mathcal{L}_{vr}$, are the mean absolute error (MAE) between predicted and target joint positions, rotations, translational velocities, and rotational velocities, respectively, in both local and world spaces. In addition to direct velocity predictions, $\mathcal{L}_{dp}$ and $\mathcal{L}_{dr}$ penalize the velocity MAE by computing the translational and rotational velocities in the local and world spaces on-the-fly via finite-difference.
This is inspired by the reconstruction loss proposed in \cite{holden2020learned} for Learned Motion Matching. 
Finally, $\mathcal{L}_{f}$ penalizes the MAE for the facing direction in the world space to prevent any character rotational drift. Loss terms are empirically weighted to be on a similar scale.
\section{Experiments and Results}
We perform a number of experiments to assess ZeroEGGS's performance regarding generalization, style control flexibility, and subjective quality. Please also refer to the supplemental video for additional visual results.
\subsection{Unseen Styles and Speakers}

\begin{figure}
     \centering
     \begin{subfigure}[b]{.475\textwidth}
         \centering
         \includegraphics[width=.4\textwidth]{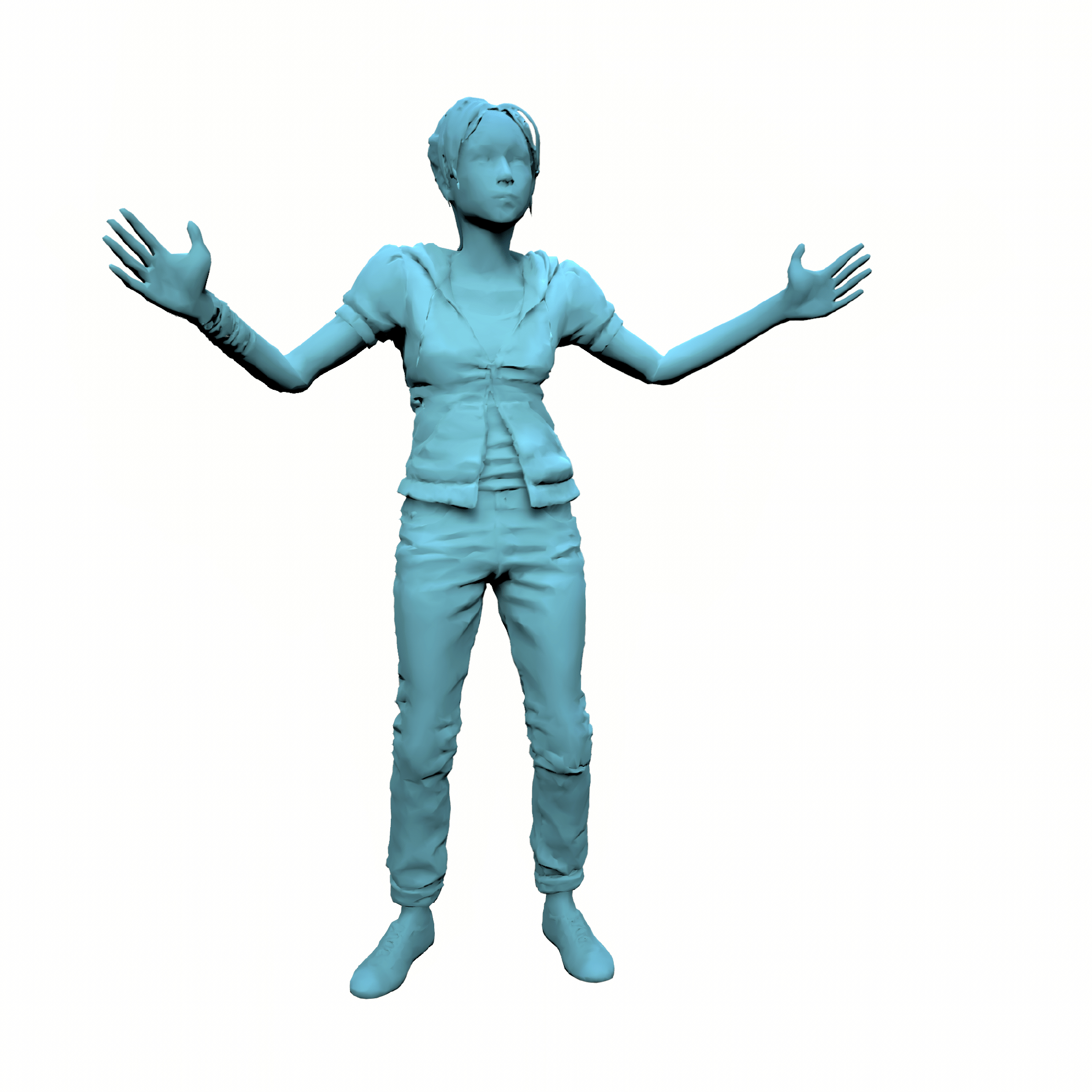}
         \includegraphics[width=.4\textwidth]{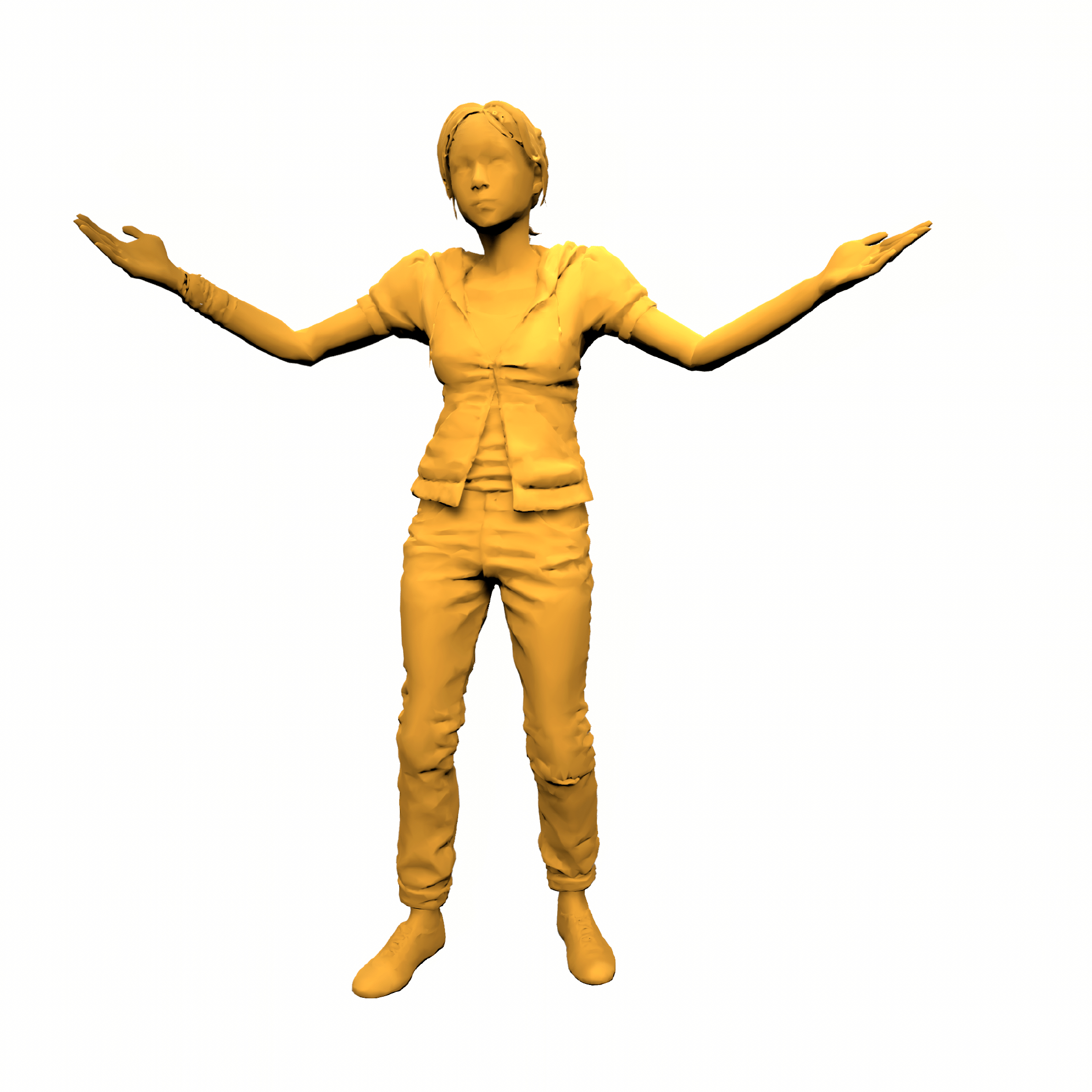}
         \caption{Oration: this style was not part of the training data in any form.}
          \label{fig:lecture_p}
     \end{subfigure}
     \vskip\baselineskip
     \begin{subfigure}[b]{.475\textwidth}
         \centering
         \includegraphics[width=.4\textwidth]{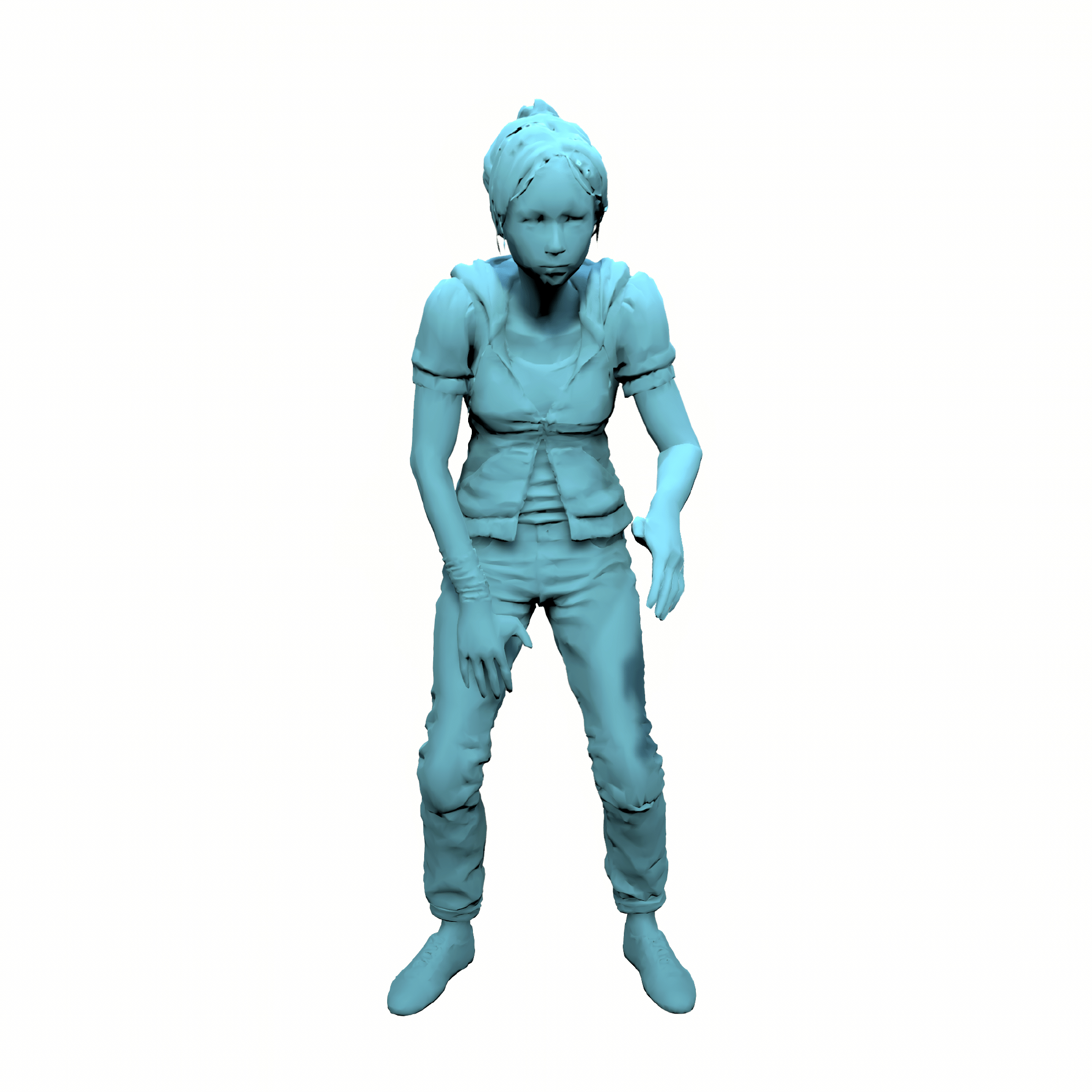}
         \includegraphics[width=.4\textwidth]{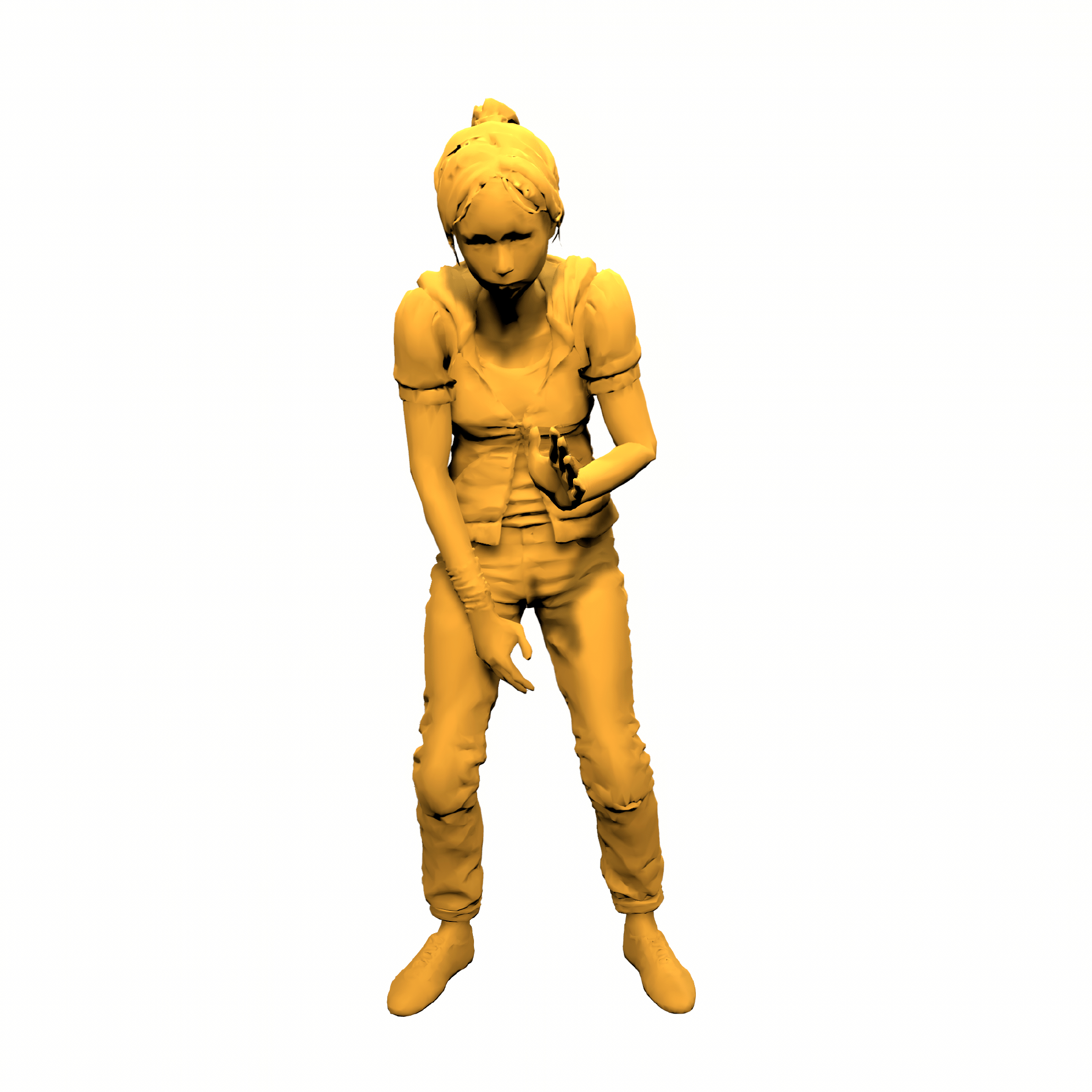}
         \caption{Tired: this style was part of the training data, but not the specific style example clip used for conditioning.}
         \label{fig:tired_p}
     \end{subfigure}
        \caption{Samples of generated gestures given unseen style examples. Left column: given example. Right column: generated gesture.}
        \label{fig:unseen_styles}
\end{figure}

With our Style Encoder, we can extract style control features from arbitrary, previously unseen gesture motion samples. To test this, we removed all \textit{Oration} style samples from the training set, a style with noticeably larger average hand height than all other styles. We visualize the model's generalizability to this unseen style in Fig.~\ref{fig:lecture_p}. We also retain one recording sample for each style and show that at inference time, the model can produce appropriate style for these samples (see Fig.~\ref{fig:tired_p}).

\begin{figure}
     \begin{subfigure}[b]{.475\textwidth}
         \centering
         \includegraphics[width=.4\textwidth]{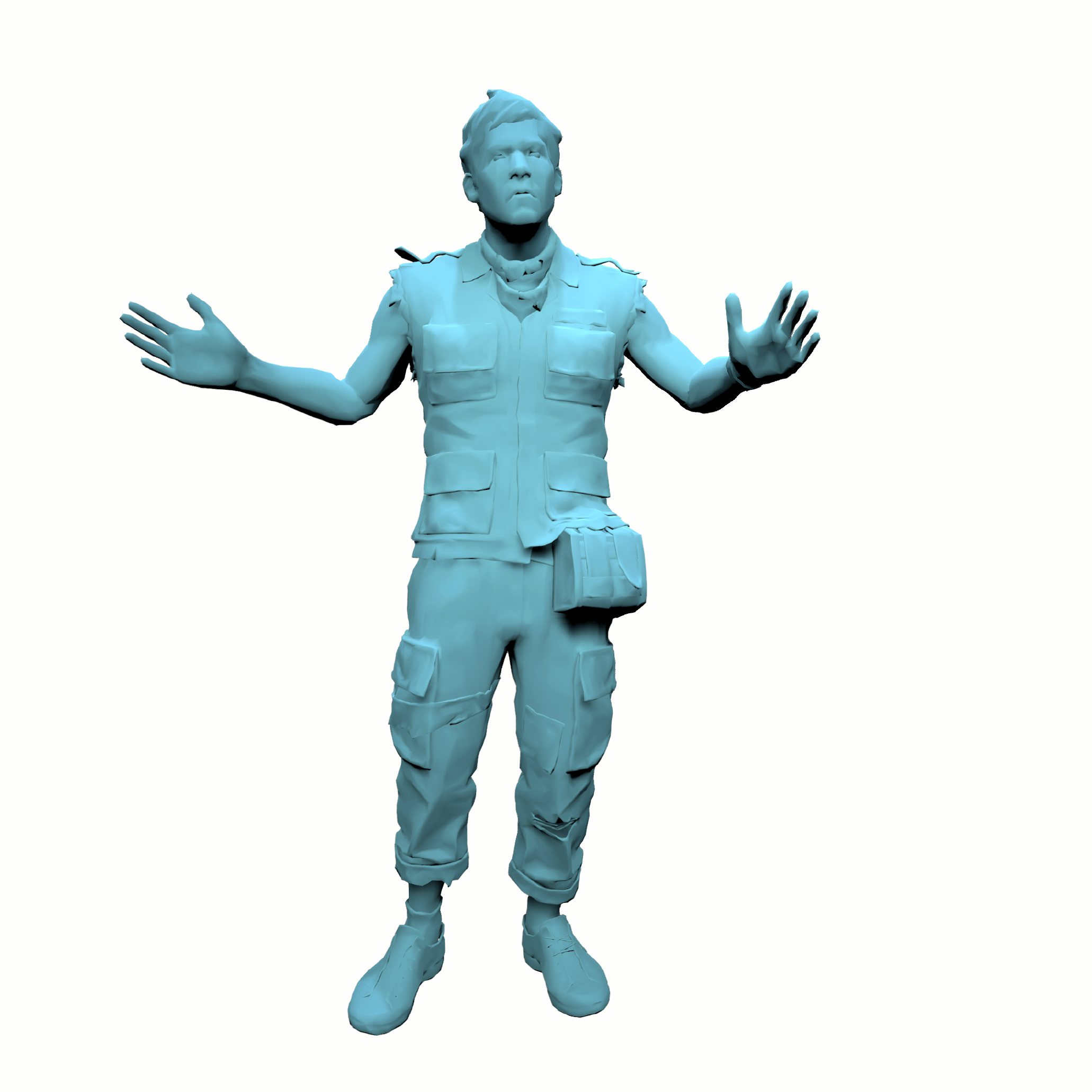}
         \includegraphics[width=.4\textwidth]{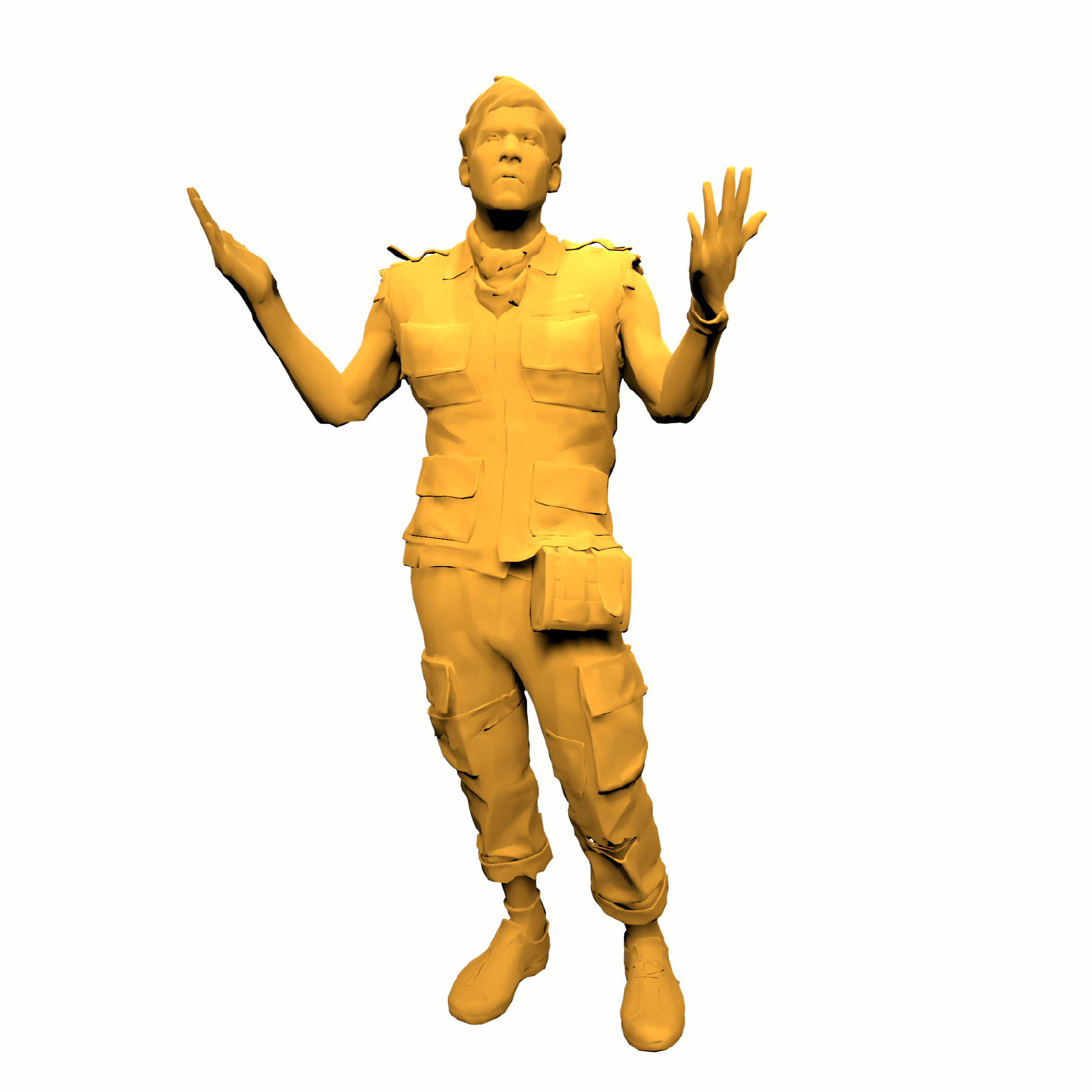}
     \end{subfigure}
        \caption{Samples of generated gestures given unseen style examples and new speaker. Left column: given example. Right column: generated gesture.}
        \label{fig:unseen_speaker}
\end{figure}

\begin{figure}
\begin{subfigure}[b]{.475\textwidth}
  \centering
  \includegraphics[width=\linewidth]{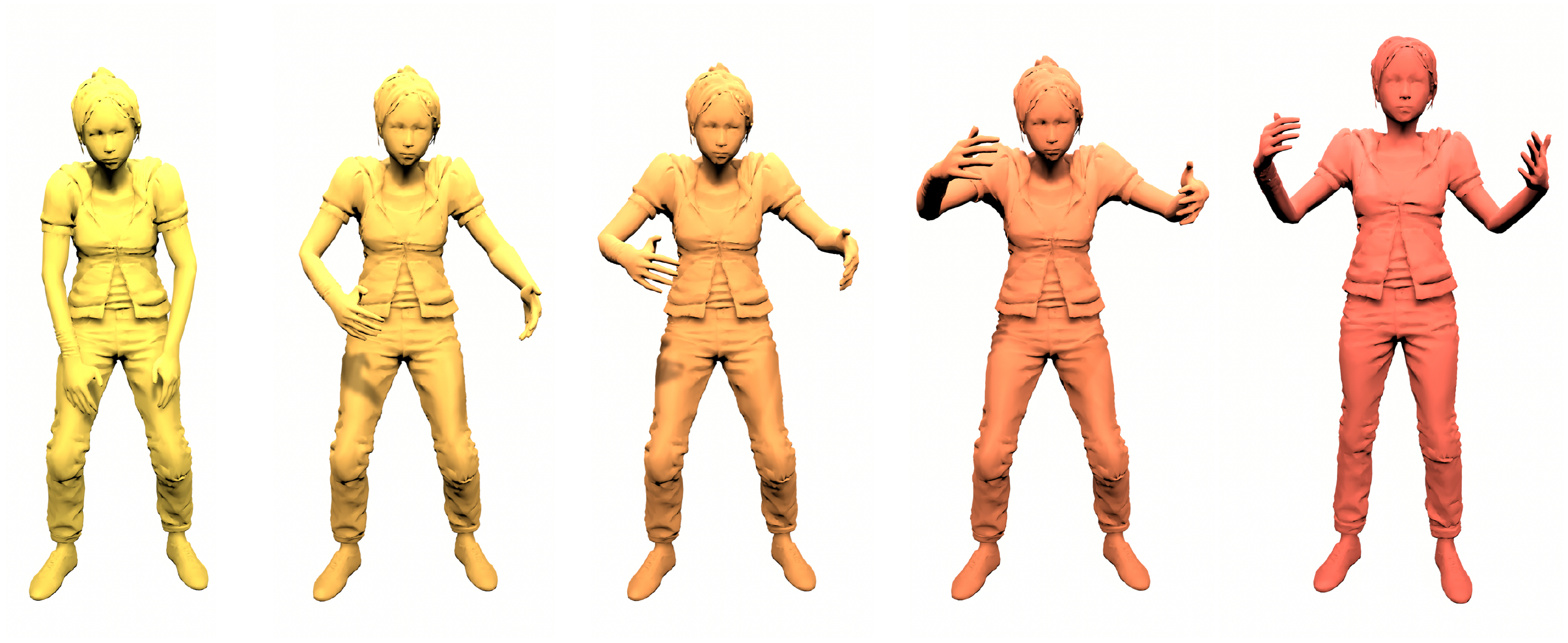}
  \end{subfigure}
  \caption{Screenshot of generated gestures by interpolating between \textit{Old} and \textit{Oration}.}
  \label{fig:interpolation_sample}
\end{figure}

Next, we assessed our model's generalization capabilities to new speakers. 
We generated results for a male speaker with a deep voice plus the unseen \textit{Oration} style. The output is visualized in Fig.~\ref{fig:unseen_speaker}. 

Because ZeroEGGS relies solely on the speech spectrum amplitude, it can be used with languages unseen in the training set. We test and confirm this language agnosticity for French, German, and gibberish, with results illustrated in the supplementary material. 
\subsection{Manipulating Style in the Style Embedding Space}
\subsubsection{Blending Styles}
The variational framework used in our Style Encoder provides a morphable and continuous style embedding space. This allows to mix the styles of multiple samples via linear interpolation. 
We illustrate interpolations of the \textit{Old} and \textit{Oration} styles in Fig.~\ref{fig:interpolation_sample}. As can be seen, the character's posture and the hand position gradually changes as we interpolate.
\subsubsection{Control via PCA Components}

\begin{figure*}[t]
     \centering
     \begin{subfigure}[b]{.39\textwidth}
         \centering
         \includegraphics[width=\textwidth]{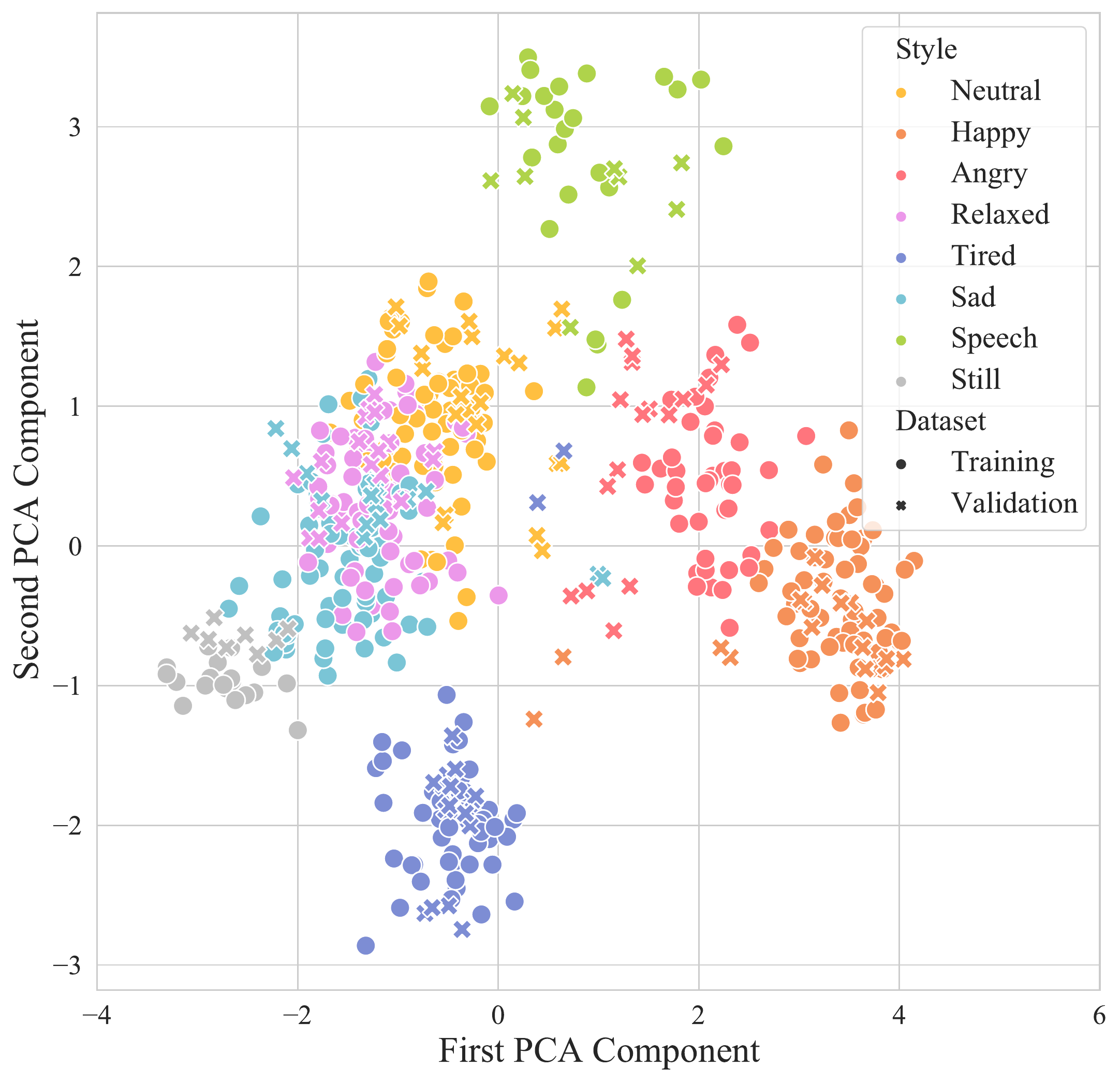}
         \caption{}
         \label{fig:PCA}
     \end{subfigure}
     \hfill
     \begin{subfigure}[b]{.181\textwidth}
         \centering
         \includegraphics[width=\textwidth]{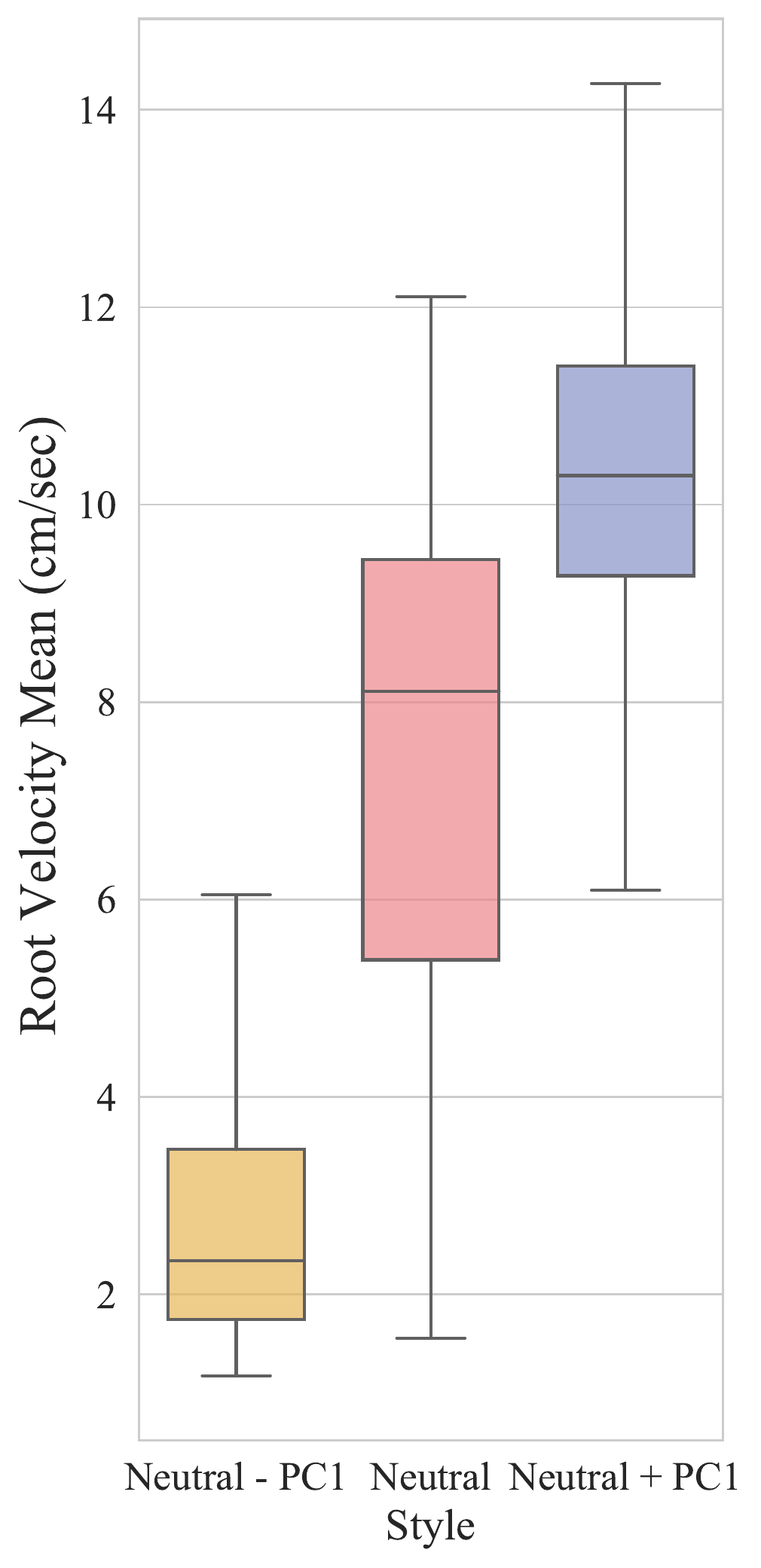}
         \caption{\vspace{0.05cm}}
         \label{fig:root_movement}
     \end{subfigure}
     \hfill
      \begin{subfigure}[b]{0.39\textwidth}
         \centering
         \includegraphics[width=\textwidth]{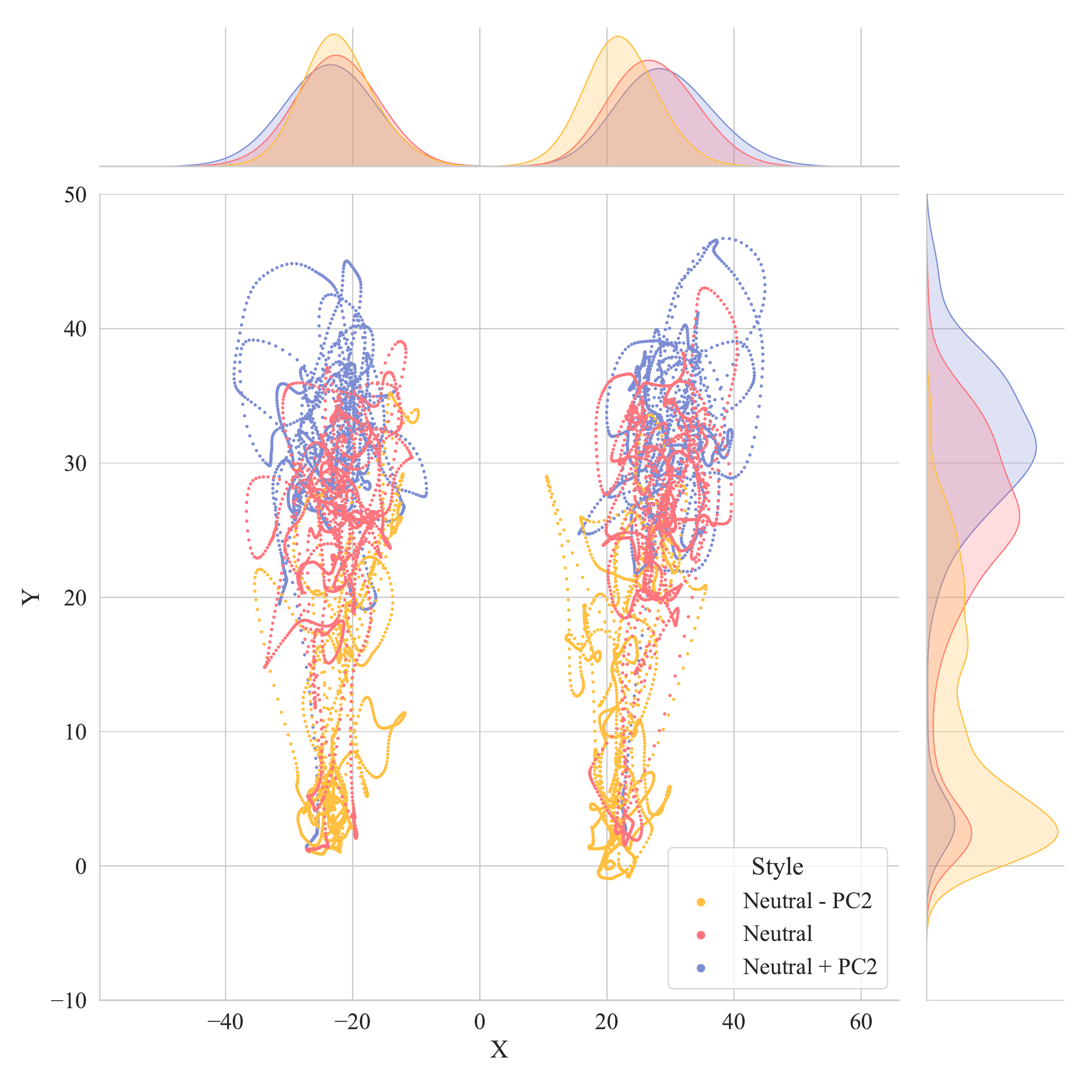}
         \caption{\vspace{.05cm}}
         \label{fig:hand_movement}
     \end{subfigure}
        \caption{We control low level gesture characteristics by projecting style embedding vector onto the PCA space. (a) Shows the first two principal components of the style embedding vector in the PCA space for a subset of styles. The first dimension corresponds to the body sway, while the second dimension corresponds to the hand height and radius. (b) We can control the root velocity of the character by changing the first principal component of the style embedding vector in the PCA space. We modify the first principal component by one standard deviation of the Neutral style. (c) Shows the hand trajectories for the neutral style after modifying the second principal component of the style embedding vector in the PCA space.}
        \label{fig:PCA_modulation}
\end{figure*}

We can control some of the low level style characteristics by projecting the style embedding vector onto a PCA space and manipulating the components. Fig. \ref{fig:PCA} shows a scatter plot of the first two principal components for non-overlapping samples in different styles. We observe that the first principal component roughly corresponds to body sway. The more static styles such as \textit{Still} and \textit{Sad} are located on the left side of the plot, while more dynamic styles, such as \textit{Happy} and \textit{Angry}, are located on the right side of the plot.

Similarly, the second principal component is associated with hand motion height and radius. For example \textit{Oration} samples, for which hands are usually above the shoulders, are located on the upper parts of the plot, whereas styles such as \textit{Tired}, during which the actor put her hands on her knees, are on the lower parts of the plot.

We can modify these gestures characteristics in the PCA space and project them back into the original style embedding space. Fig. \ref{fig:root_movement} shows the distribution of the root velocity as an indicator of body sway for three variation a generated gesture: A \textit{Neutral} style example and its two versions obtained by changing its first principal component. We can see that modifying the first principal component affects root velocity. Fig. \ref{fig:hand_movement} shows the hand movement trajectories relative to the hip joint in a frontal view for the same \textit{Neutral} sample, after manipulation of the second principal component. The distribution plot on the Y axis of the Figure highlights the effect of this component on hand height.

\subsection{User study}
We evaluated subjective result quality in three separate user studies, assessing the motion output with regards to (1) naturalness, (2) appropriateness w.r.t. the speech, and (3) recognizability of style. We compared performance of our model to the ground truth reference motion as well as three baselines, as described below.
\subsubsection{Study Design}\label{sec::study_design}
We used a MUSHRA-like (MUltiple Stimuli with Hidden
Reference and Anchor) \cite{ITU2015} interface based on \cite{jonell_2021_hemvip}. We had a total of 131 participants, with a minimum of 20 per study (ages 20-55 years  $\mu$ = 33.6, $\sigma$ = 8.1). 
In each study, participants completed 12 pages of ratings, on each of which they compared the motion of rendered animation clips with the female character shown in Fig.~\ref{fig:unseen_styles}, for the four systems listed further below. The order of systems and rating pages was randomized. For \textit{naturalness}, participants were asked how natural the character's motion looks to them. For \textit{appropriateness}, participants rated how appropriate the gesture motion was for the speech. For \textit{style portrayal}, videos played without sound and participants judged how well the motion represented a given style. For this, a fixed set of 6 representative styles (\textit{Happy}, \textit{Sad}, \textit{Relaxed}, \textit{Old}, \textit{Angry}, \textit{Speech}), and participants rated two samples of each. 

Participants compared and rated the following 4 systems: 
\begin{enumerate}
  \item Ground truth motion (GT)
  \item Mismatched motion (MM)
  \item Our model (ZEGGS)
  \item MoGlow (MG) 
\end{enumerate}

GT is the original motion-captured animation associated with the given speech segment.  MM is also motion-captured animation but used in the wrong context. For the naturalness and appropriateness studies, MM is a motion originally associated with a \textit{different} speech segment of the same style, while in the style portrayal study, MM is a motion associated with a different style.

As a strong baseline model, we chose MoGlow (MG) by Alexanderson et al.~\cite{alexanderson2020style} which represents style via their three proposed gesture features: hand height, velocity and radius. We chose this model based on its style-control capabilities, competitive performance \cite{kucherenko2021large}, and code availability. We compare model statistics of ZEGGS and MG in Table~\ref{tab::stats}.
We first retrained the model using our dataset, however, this led to severely jittery, low quality output. We therefore decided to train MoGlow using its original training dataset, the Trinity Speech-Gesture dataset I (TSG)~\cite{ferstl2018investigating}. TSG contains 4 hours of a single male speaker producing spontaneous speech in monologue format in an overall neutral style. 
To level the playing field, we therefore ran 2 versions of each of the three studies of naturalness, appropriateness, and style: The first version used our dataset, the second version used the TSG dataset. For the TSG version of the style study, the style input motion features were taken from our own dataset since TSG does not contain any style data.
A number of video samples from our user study are included in the supplemental material.


\subsubsection{Results}

\begin{figure}[t]
  \centering
  \includegraphics[width=0.7\linewidth]{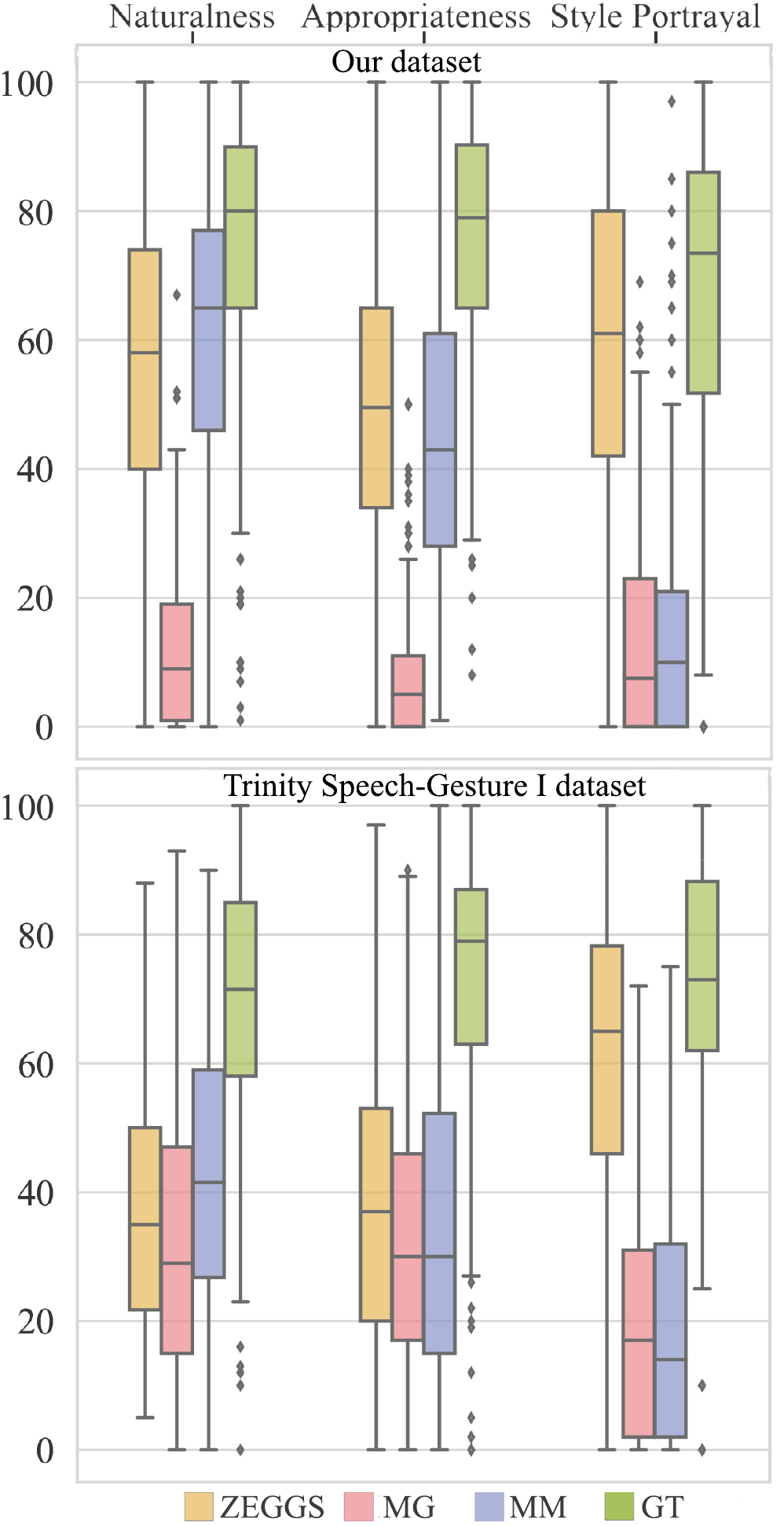}
  \caption{Results from our three user studies. The x-axis represents the system and the y-axis the average participant ratings.  }
  \label{fig:boxplot}
\end{figure}

We analyzed the collected perceptual data using Analysis of Variance (ANOVA) when the normality assumption was not violated (Shapiro–Wilk test) and corrected degrees of freedom using the Greenhouse–Geisser when sphericity was violated. When normality was violated, we used Aligned Rank Transform (ART) instead of ANOVA. We performed post-hoc comparisons using Estimated Marginal Means. We had one within-subject factor for naturalness and appropriateness measures, the gesture generation system, with 4 levels: GT, MM, ZEGGS, MG. For style portrayal, we additionally had the factor style. Results are visualized in Fig.~\ref{fig:boxplot} and summarized in Table~\ref{tab::stats}.


The results of our user studies show that our model outperforms the comparison model \cite{alexanderson2020style} for all three collected measures, motion naturalness, appropriateness for speech, and recognizability of portrayed style for our own dataset, and for appropriateness and style recognizability for the TSG data. There was a main effect of system for each of the three measures, and GT performed best across each measure, as expected ($p<.001$ for all pairwise comparisons with GT).

For \textbf{naturalness}, we would expect a similar level of naturalness for MM as for GT because they are both motion-capture clip. However, the mismatch with the speech audio resulted in lower ratings which highlights the importance of synchronicity for the perception of naturalness. MM was rated significantly lower than GT and higher than MG (for TSG, $p<.05$ for ZEGGS vs. MM, all other differences $p<.001$). For our dataset, MM was not rated significantly different to ZEGGS. 

For \textbf{appropriateness}, for our own dataset, ZEGGS scored higher than MM ($p<.01$), indicating that our model was indeed able to capture some relation between the speech and motion modalities and produced output that was more appropriate for the given speech than the baseline systems. This reinforces the idea that synchronicity plays an important role in the perception of gesture quality. Here, even if MM is a motion-captured clip in the same style as the target, it seems inappropriate for the speech segment, because gesture are out of sync with voice inflections. However, for the TSG dataset, MM was rated on-par with ZEGGS and MG.

\begin{figure}[t]
  \centering
  \includegraphics[width=1.0\linewidth]{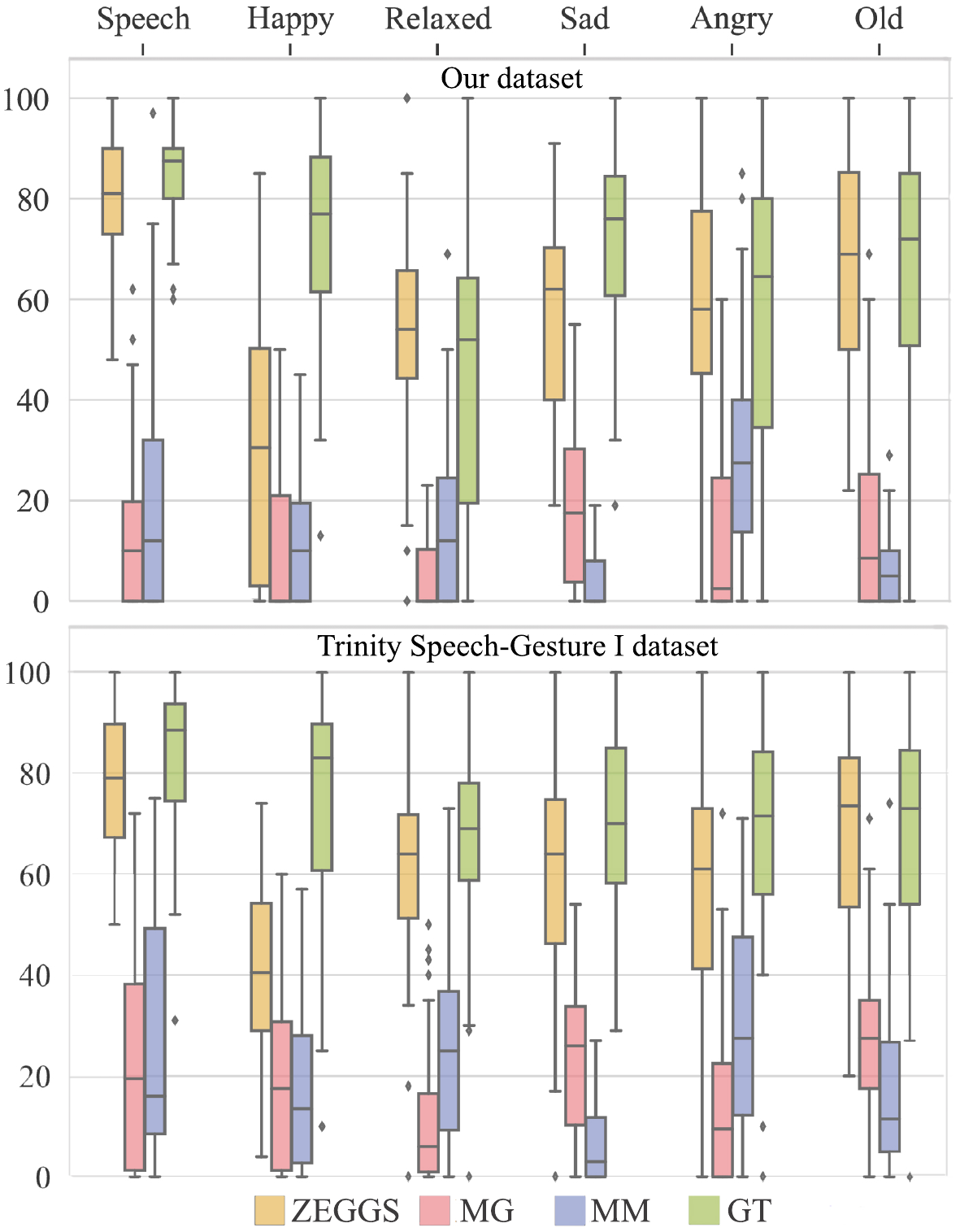}
  \caption{Rating scores per style for the style portrayal study.}
  \label{fig:interaction}
\end{figure}

When measuring \textbf{style portrayal} specifically, there is also a significant interaction between the factors system and style. This time, MM is a motion-captured animation from another style. Therefore it performs significantly worse than GT as it accurately portrays a different style ($p<.001$). MM was also rated significantly worse than ZEGGS ($p<.001$) but not MG, indicating that MG was unable to portray the desired styles whereas our method produced good results. Visually inspecting the interaction of system and style, this appears mainly stem from a lower than average performance of ZEGGS for the style \textit{Happy} (see Fig.~\ref{fig:interaction}).

We believe there are several reasons explaining why ZeroEGGS outperformed MoGlow in most subjective evaluations. 
Firstly, as described in Sec.~\ref{sec::study_design}, training MoGlow on our own dataset yielded poor results and we therefore trained it on the single-style TSG dataset with which it was originally developed. 
This means that relatively lower performance of MG on our dataset compared to on TSG is to be expected and can be seen in Fig.~\ref{fig:boxplot} (compare top to bottom).
While for ZEGGS, there is also a performance drop on TSG, compared to on our own dataset, ZEGGS still performs overall better than the MG model that was trained on this dataset. 
Regarding naturalness, animation produced by MoGlow was more jittery and the lower generated frame rate (20 fps instead of 60 fps) may have also played a role in the evaluation. 
For style portrayal, the style representation space that ZeroEGGS uses captures a lot of information that is discarded by the 3 features used by MoGlow. By collapsing style information to a few hand movement statistics, MG cannot for example accurately capture style differences of gestures with the general same movement intensity, but different trajectories. This impedes learning from a dataset containing numerous very different styles. Moreover, the 3 MG style features only describe hand movement, which means that the lower half of the body is not stylized. 
The absolute value of the score is not indicative of the absolute performance of the model, but rather an indication of the models' relative performance magnified by the nature of MUSHRA-like tests.

\begin{table*}[]
    \centering
	\caption{Comparison of our model (ZEGGS) with the baseline model (MG). }
    \begin{tabular}{||c|c|c|ccc|ccc||}
    \hline
    \multirow{2}{*}{System} & \multirow{2}{*}{\#Parameters} & {Inference time} & \multicolumn{3}{c|}{Rating on our dataset} & \multicolumn{3}{c||}{Rating on TSG}\\ 
    & & (per frame) & Nat. & Appr. & Style & Nat. & Appr. & Style \\ \hline  \hline
    MG & 86M (88M with style) & 29ms & 11.8$\pm$ 12.2 & 7.8$\pm$ 9.6 & 12.9$\pm$ 16.1 & 32.7$\pm$ 21.0 & 32.2$\pm$ 20.6 & 19.8$\pm$ 18.0\\ \hline
    ZEGGS & 25M & 4ms & 56.3$\pm$ 22.2 & 48.9$\pm$ 20.3 & 58.1$\pm$ 25.9 & 36.7$\pm$ 18.4 & 38.3$\pm$ 23.3 & 61.9$\pm$ 23.3 \\ \hline

    \end{tabular}
    \label{tab::stats}
\end{table*}

\section{Discussion}
Our experiments show that our model generalizes to new voices, languages and to new styles that were not part of the training data. This zero-shot style transfer capability constitutes a major advantage when compared to existing methods. In practice, it means that a single model can be used for different characters and that it does not have to be retrained each time a new style is needed. Moreover, specifying style directly using examples enables the model to capture smaller details that could be otherwise discarded by the strict information bottleneck imposed by handcrafted features and statistics. This results in more realistic and compelling motions. Finally, by using examples instead of labels, users do not have to elaborate a precise consentaneous style taxonomy. 

We also show that our variational framework learns a meaningful style embedding that enables manipulation and interpolation within the latent space. This allows, for example, the weighted mixing of style reference samples. This proves useful in practice when transitioning from one style to another in the same animated sequence. In addition, the probabilistic nature of variational methods allows for variations of gesture motion for a given input by re-sampling.

In our subjective evaluations, ZeroEGGS outperformed the baseline model w.r.t. naturalness, appropriateness for speech, and style portrayal. This means that for ZeroEGGS, generalization does not come at the cost of visual quality and style diversity. Improvement on state-of-the-art mainly comes from the superior descriptive power of our example-based conditioning over handcrafted features and statistics. As explained above, capturing more detail in the style embedding improves naturalness, but it also helps generating subtleties that helps differentiate between styles. Moreover, our loss ensures smooth and stable motion, which also plays a role in the perception of the naturalness of the animation.

While our model outperforms our baseline models, it does fall significantly behind ground truth motion, motivating further work to improve the naturalness of the generated output. Moreover, some more subtle motions that are specific to certain styles are often over-smoothed or even not produced. For instance, in the \textit{Agreement} style, head nods are very infrequently generated, while abundantly present in the training data. Enforcing explicit disentanglement in the style latent space could help capture motion localized in specific parts of the body. 
Our model showed a slight decrease in performance on the Trinity Speech-Gesture dataset, on which it was not trained, likely due to the frequent displacement of the speaker, which is not captured in our dataset. Our method performs well on more static style, but we would like to address character displacement as a next step.

Finally, ZeroEGGS generates small amounts of foot sliding, a common problem in non physically-based animation generation that can be addressed with IK-based post-processing.

ZeroEGGS uses raw spectral audio features as input which provides robustness to language and speaker voice. However, this means that the generation is limited to so-called ``beat gestures'' that do not carry semantic information about the speech content. This explains in part why the ground truth motion scored significantly higher w.r.t. appropriateness for the speech in the user study. Future work could include semantic markers in the speech as input to the model.
\section{Conclusion}
We proposed ZeroEGGS, a model that generates stylized gesture animations from speech. The desired style is specified to the model as a short example clip. The model generalizes beyond training data allowing the generation of unseen styles, for different voices and languages. Moreover, the latent style representation allows for control over generation. Our experiments show that ZeroEGGS generates state-of-the-art gesture animations.

There are many directions for future work, including enforcing user-specified disentanglement in the learned style latent space, providing support for semantic gestures as well as exploring new architectures based on affine transform coupling layers.
\bibliographystyle{eg-alpha-doi}  
\bibliography{ZEGGS}        


\end{document}